\documentclass[a4paper,preprint,amsmath,showpacs,amssymb,floatfix,prl,preprintnumbers,footinbib]{revtex4}
\usepackage[dvips]{graphicx}
\usepackage{dcolumn}
\usepackage{bm}

\begin{document}

\newcommand{\comment}[1]{}
\newcommand{\T}{\mathcal{T}}


\title{Rashba effect induced localization in quantum networks} 

\author{Dario Bercioux$^1$, Michele Governale$^2$, Vittorio
Cataudella$^1$, Vincenzo Marigliano Ramaglia$^1$} 
\affiliation{$^1$Coherentia-INFM and Dipartimento di Scienze Fisiche
Universit\`a degli studi ``Federico II'', Napoli, Italy\\
$^2$NEST-INFM and Scuola Normale Superiore, Piazza dei Cavalieri 7, 
I-56126 Pisa, Italy}

\date{\today}

\begin{abstract}
We study a quantum network extending in one-dimension (chain of square
loops connected at one vertex) made up of quantum wires with Rashba
spin-orbit coupling.  We show that the Rashba effect may give rise to
an electron localization phenomenon similar to the one induced by
magnetic field.  This localization effect can be attributed to the spin
precession due to the Rashba effect.  We present results both for the
spectral properties of the infinite chain, and for linear transport
through a finite-size chain connected to leads.  Furthermore, we study
the effect of disorder on the transport properties of this network.
\end{abstract} 

\maketitle

\textit{Introduction.}
It has been recently shown that in a particular class of
two-dimensional lattices quantum interference due to the the
Aharonov-Bohm effect and to the geometry of the network can induce
strong electron localization~\cite{vidal-1998,vidal-cage-2000}.  In
such systems when localization occurs particle motion is confined by
destructive interference inside a small portion of the network which
is called Aharonov-Bohm (AB) cage.  This kind of localization does not
rely on disorder\cite{anderson-1958} but only on quantum-interference
and on the geometry of the lattice.  There have been several
theoretical works addressing different aspects of AB cages as the
effect of disorder and electron-electron interaction
\cite{vidal-prb-long},
interaction induced delocalization\cite{vidal-cage-2000}, and
transport \cite{vidal-tra-2000}.  From the experimental side, the
AB-cage effect has been demonstrated for
superconducting\cite{abilio-1999} and metallic
networks\cite{naud-2001} in the so called $\mathcal{T}_3$ lattice.
    
As already stated before, in the AB cages localization is due to
interference stemming from the fact that an electron traveling along
different paths acquire different phases.  
It is known that the
wavefunction of an electron moving in the presence of Spin-Orbit (SO)
coupling acquires quantum phases due to the Aharonov-Casher effect
\cite{casher,mathur,
balatsky,aronov-1993,splettstoesser-2003,frustaglia}. We now focus on
the Rashba SO coupling
\cite{rashba-1960}, which is present in semiconductor 
heterostructures due to lack of inversion symmetry in growth
direction.  It is usually important in small-gap zinc--blende--type
semiconductors, and its strength can be tuned by external gate
voltages, as it has been demonstrated
experimentally\cite{nitta-1997,schaepers-1998,grundler-2000}.
 
\begin{figure}
	\centering \includegraphics[width=3.0in]{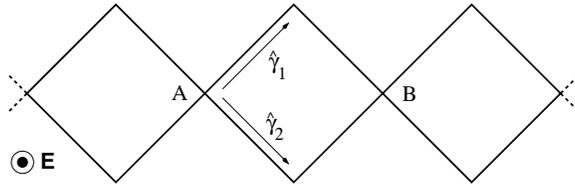}
	\caption{Schematic view of the diamond chain. The bonds are
	single-channel quantum wires with SO coupling. In the ideal
	case all bonds have the same length $L$. The unitary cell
	contain three nodes (4 wires): one with coordination number 4
	and two with coordination number 2.  \label{fig_graph}}
\end{figure}

The question we address in this Letter is whether it is possible to
have localization in quantum lattices induced only by the SO coupling
without magnetic fields.  To answer this question we study the minimal
model of a bipartite structure containing nodes with different
coordination numbers that with magnetic field exhibits electron
localization.  This model structure is a linear chain of square loops
connected at one vertex (see Fig.~\ref{fig_graph}), which we term
\textit{diamond chain}.  We have in mind a realization in a
semiconductor heterostructure where the bonds are single-channel
quantum wires of length $L$ with Rashba SO coupling. External gates
should be present to tune the strength of the SO coupling.  This
one-dimensional lattice retains the essential features of the more
complex $\mathcal{T}_3$ networks, allowing for simple (even analytical
for the spectrum) solutions.
%
%
%
%

\textit{Model and formalism.}
Neglecting subband hybridization due to the Rashba
effect\cite{mireles-2001, governale-2002}, the Hamiltonian for a
single-channel wire along a generic direction $\hat{\gamma}$ in the
$x$--$y$ plane reads
%
%
\begin{equation}
\label{hamiltonian}
	\mathcal{H} = \frac{p_{\gamma}^2}{2\,m} -
	\frac{\hbar k_{\text{SO}}}{m}p_\gamma\left(
\vec{\sigma} \times \hat{z}\right)\cdot\hat{\gamma},
\end{equation}
%
%
where $k_{\text{SO}}$ is the SO coupling strength, and $\vec{\sigma}$
the vector of the Pauli matrices.  The SO coupling strength
$k_{\text{SO}}$ is related to the spin precession length
$L_{\text{SO}}$ by $L_{\text{SO}}=\pi/k_{\text{SO}}$.  For InAs
quantum wells the spin-precession length ranges from $0.2$ to $1$
$\mu$m~\cite{nitta-1997,schaepers-1998,grundler-2000}. These are the
characteristic length scales required for the bonds of the network for
spin precession to be effective.  In order to calculate spectral and
transport properties of the network we need to write the wavefunction
on a bond (quantum wire) connecting the nodes $\alpha$ and $\beta$,
along the direction $\hat{\gamma}_{\alpha\beta}$
%
%
\begin{eqnarray}
\label{wavefunction}
	\mathbf{\Psi}_{\alpha\beta}(r)&=&\frac{1}{\sin(k
	  l_{\alpha\beta})} \left\{ \sin\left[ k
	  (l_{\alpha\beta}-r)\right]
	  e^{i(\vec{\sigma}\times\hat{z})\cdot
	  \hat{\gamma}_{\alpha\beta}~ k_{\text{SO}} r}
	  \mathbf{\Psi}_{\alpha} \right. \nonumber \\ && \left. +
	  \sin(k r) e^{i(\vec{\sigma}\times\hat{z})\cdot
	  \hat{\gamma}_{\alpha\beta} ~ k_{\text{SO}}
	  (r-l_{\alpha\beta})} \mathbf{\Psi}_{\beta} \right\}
\end{eqnarray}
%
%
where $k$ is related to the eigen energy by
$\epsilon=\frac{\hbar^2}{2m} (k^2-k_{\text{SO}}^2)$\cite{note2}, $r$
is the coordinate along the bond, and $l_{\alpha\beta}$ the length of
the bond.  The spinors $\mathbf{\Psi}_{\alpha}$ and
$\mathbf{\Psi}_{\beta}$ are the values of the wavefunction at the
nodes $\alpha$ and $\beta$ respectively.  The  spin
precession due to the Rashba effect is described by the exponentials
containing Pauli matrices in Eq.~(\ref{wavefunction}).

Eq.~(\ref{wavefunction}) is the key step to generalize the existing
methods to study quantum networks \cite{kottos-1999,vidal-tra-2000} in
the presence of Rashba SO coupling.  The wavefunction of the whole
network is obtained by imposing the continuity of probability current
at the nodes. For a generic node $\alpha$ it reads:
%
%
\begin{equation}\label{continuity}
	\textbf{M}_{\alpha\alpha} \mathbf{\Psi}_\alpha+\sum_{\langle
	\alpha,\beta\rangle}
	\textbf{M}_{\alpha\beta}\mathbf{\Psi}_\beta=0,
\end{equation}  
%
%
where
%
%
\begin{subequations} \label{ms}
\begin{eqnarray} 
	\textbf{M}_{\alpha\alpha} &= & \sum_{\langle
	\alpha,\beta\rangle} \cot k l_{\alpha\beta}\\
	\textbf{M}_{\alpha\beta} &= & -
	\frac{\exp\left[-i(\vec{\sigma}\times\hat{z})\cdot
	\hat{\gamma}_{\alpha\beta}~
	k_{\text{SO}}l_{\alpha\beta}\right]}{ \sin k l_{\alpha\beta}}.
\end{eqnarray}
\end{subequations}
%
%
In  Eqs.~(\ref{continuity},\ref{ms}) the 
sum $\sum_{\langle \alpha,\beta\rangle}$  runs over all 
nodes $\beta$ which are connected by a bond to the node $\alpha$. 
%
%
\begin{figure}
	\centering
	\includegraphics[width=3.in]{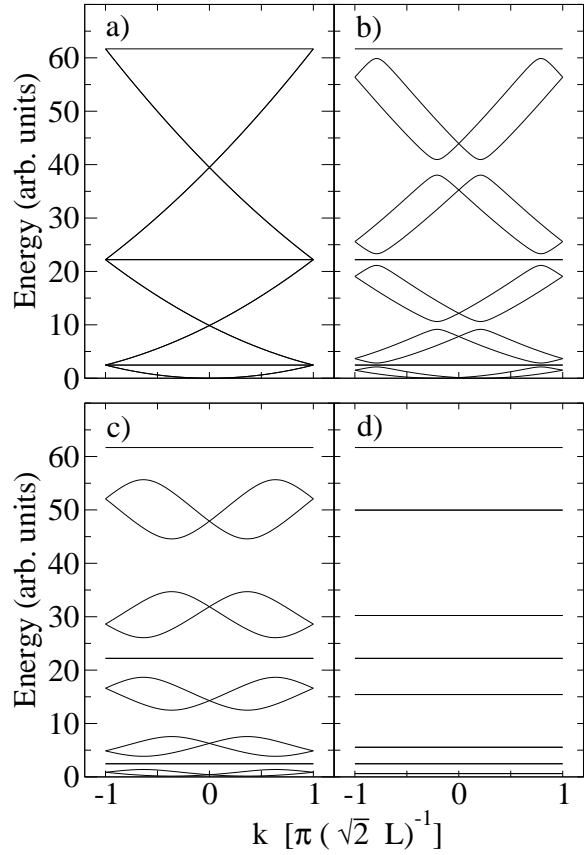}
	\caption{Spectrum of the diamond chain for different values of
	the strength of the spin-orbit coupling: a)
	$k_{\text{SO}}L=0$; b) $k_{\text{SO}}L=0.5$; c)
	$k_\text{{SO}}L=1.0$; and d) $k_\text{{SO}}L=\pi/2$.
\label{spectrum}}
\end{figure}
%
%

\textit{Spectral properties.}
Now, we apply the method presented above to calculate the spectral
properties of the diamond lattice.  For an infinite lattice this can
be done imposing the Bloch condition on the wavefuction in the unit
cell.  This straightforward procedure yields for the spectrum the
following analytical expressions
%
%
\begin{small}
\begin{eqnarray}\label{spectrum-a}
        \varepsilon^{(0)}_n(k)&=&\left( \frac{\pi }{2} + n\,\pi
        \right)^2 \\ \nonumber \varepsilon_n^{(\pm)}(k)&=&\left\{ n
        \pi + \arccos\left[ \frac{1}{2}\left(2 + 2\cos ({\sqrt{2}}k L)
        \cos (k_{\text{SO}} L)^2\right.\right.\right.\\ &&
        \left.\left.\left. \pm \sqrt{2}\sin (\sqrt{2}kL)\sin
        (2k_{\text{SO}}L)\right)^{\frac{1}{2}} \right]
        \right\}^2. \label{spectrum-b}
\end{eqnarray}
\end{small}
%
%
The momentum $k$ is defined in the first Brillouin zone
$\left[-\frac{\pi}{\sqrt{2}\,L},\frac{\pi}{\sqrt{2}\,L} \right]$
(notice that the lattice constant is $\sqrt{2}\,L$).  The spectrum is
composed by three kinds of bands. The first one is non dispersive:
this is a characteristic of every bipartite structures containing
nodes with different coordination numbers.  The bands $\pm$ are
degenerate for zero SO coupling, and are split by it. From
Eq.~(\ref{spectrum-b}) it is apparent that these bands become
nondispersive for $k_{\text{SO}}L = (n+\frac{1}{2}) \pi$, being $n$ an
integer. This condition can be recast using the spin-precession length
as $L= (n+\frac{1}{2}) L_{\text{SO}}$.  For these value of the SO
coupling strength the system becomes localized (as it is indicated by
the diverging effective mass).  A portion of the spectrum
Eqs. (\ref{spectrum-a},
\ref{spectrum-b}) is shown in Fig. (\ref{spectrum}) for increasing
values of the SO coupling strength.  For zero SO coupling
there are no gaps in the spectrum.  For finite values of the Rashba
coupling the spin degeneracy of the $\pm$ bands is lifted and gaps
open in the spectrum.  When the SO coupling strength approaches the
value $k_{\text{SO}}L=\pi/2$ the spectrum collapses to a series of non
dispersive bands.

The localization for the diamond chain can be understood in terms of
interference effects in analogy to the AB cages.  In the case of the
AB effect the phase difference between different paths is due to the
enclosed magnetic flux.  In the present case the Rashba effect is
responsible for it.  Consider an electron with spin $|\sigma\rangle$
in A (see Fig.~\ref{fig_graph}).  It can reach point B either via the
upper or the lower path.  When traveling along the upper path, the
spin undergoes a precession first around $\hat{z}\times
\hat{\gamma}_1$ and then around $\hat{z}\times \hat{\gamma}_2$. Hence,
the final state in B is given by
$\mathcal{R}_{\hat{\gamma}_2}\mathcal{R}_{\hat{\gamma}_1}|\sigma\rangle$,
where $\mathcal{R}_{\hat{\gamma}}=\exp\left[-i
\vec{\sigma}\cdot(\hat{z}\times\hat{\gamma}) k_{\text{SO}}L\right]$.
Similarly for propagation along the lower path, the state in B is
$\mathcal{R}_{\hat{\gamma}_1}\mathcal{R}_{\hat{\gamma}_2}
|\sigma\rangle$.  Destructive interference occurs when
$\{\mathcal{R}_{\hat{\gamma}_1},\mathcal{R}_{\hat{\gamma}_2}\}=0$,
being $\{\ldots\}$ the anticommutator. For our setup with
$\hat\gamma_1\cdot\hat\gamma_2=0$ this condition is fulfilled if
$k_{\text{SO}}L=(n+1/2)\pi$.  A similar analysis can be carried out
for more complex structures.  In particular, it can be shown that there
are bipartite linear chains with a more complex unit cell than the
diamond chain that exhibit localization.  In analogy to the AB cages,
we call the elementary square loop in our structure a \textit{Rashba
cage}.

\begin{figure}
	\centering \includegraphics[width=3.in]{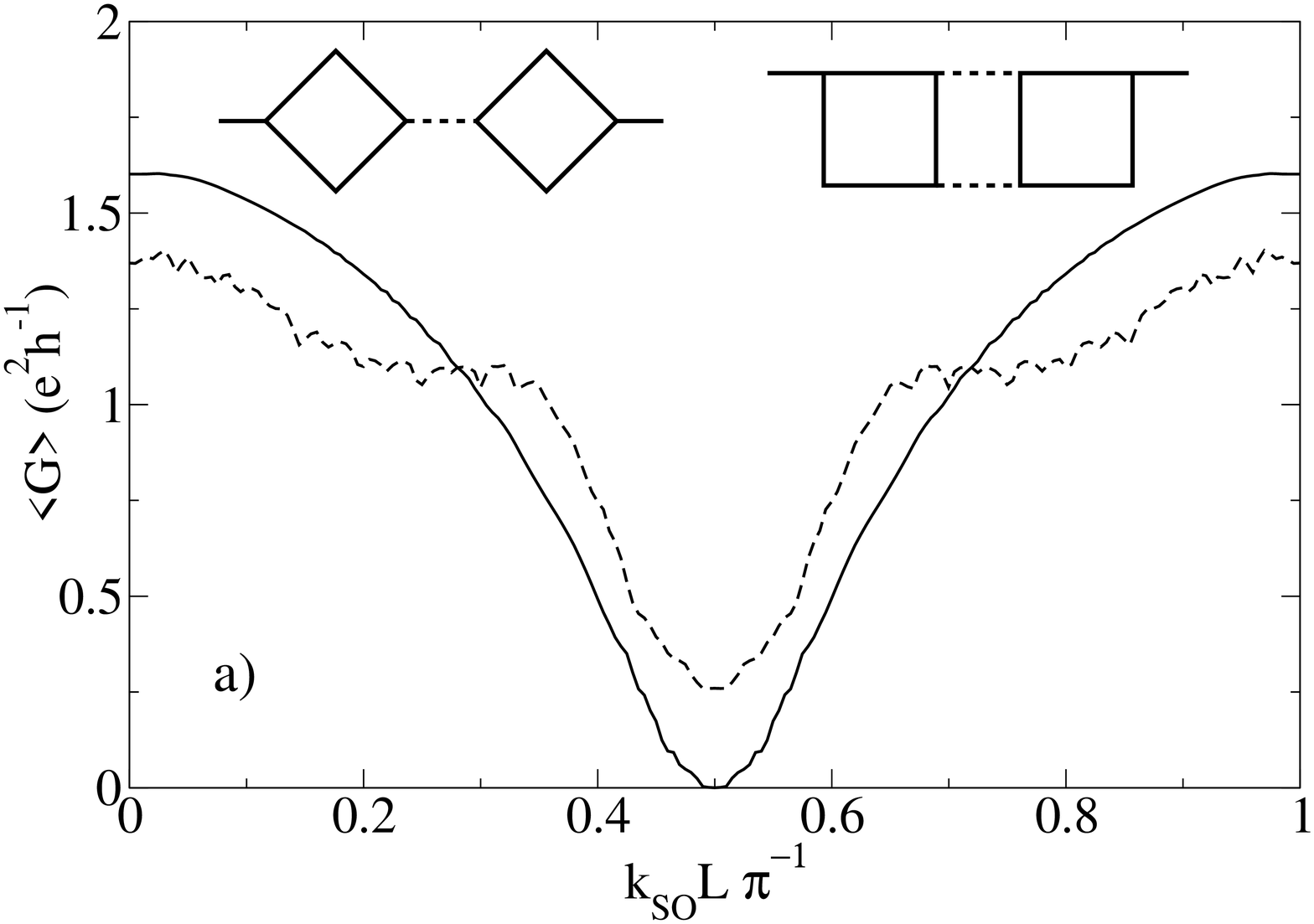}
	\includegraphics[width=3.in]{figure4.eps} 
	\caption{Panel a): Conductance (averaged over $k_{\text{in}}$
	as a function of the spin-orbit coupling strength for the
	diamond chain (continuous line) and for the 
        ladder 
	(dashed line).  The two finite-size  
         systems
        connected to
	input/output leads are shown in the inset. The parameters used
	for the calculation are: 50 elementary loops, $k_{\text{in}}$
	uniformly distributed in $[0, \pi/L]$.\\ Panel b): Conductance
	as a function of the spin-orbit coupling strength for the
	diamond chain (continuous line) and for the 
        ladder 
	(dashed line) for a fixed value of
	$k_{\text{in}}=k_{\text{F}}$. The parameters used for the
	calculation are: 50 elementary loops, $k_{\text{F}} L=n \pi +
	2$, being $n$ an integer.
\label{con_kf}}
\end{figure}

\textit{Transport properties: clean case.}
In experiments the onset of localization in a quantum network is
usually detected by transport measurements. For example, for the AB
cages the conductance is suppressed for special values of the magnetic
field.  To propose a possible experimental verifications of the
Rashba-cage effect we now evaluate the linear conductance for a
diamond chain of finite length. Furthermore, to show that this
localization effect is due to the peculiar geometry of the lattice
(bipartite containing nodes with different coordination numbers), we
contrast the diamond chain with square ladder, i.e.  a chain of square
loops connected at two vertices, (see inset of Fig.~\ref{con_kf}). In
the following, we will also refer to the latter geometry simply as
ladder.  We evaluate the conductance making use of the
Landauer--B\"uttiker formalism
\cite{landauer-1988,buttiker-1988}.   
We consider a finite piece of lattice connected to semi-infinite leads
(with no SO coupling) modeling reservoirs (see inset of
Fig.~\ref{con_kf}). To compute the transmission coefficients we
proceed along the lines of Ref.~\cite{vidal-tra-2000}.  We inject from
the left wire an electron with spin $\sigma=\pm$ along a generic
direction, whose corresponding spinors are $\mathbf{\chi}_{\sigma}$.
The wavefunctions on the external leads are simply
%
%
\begin{eqnarray}
	\Psi_{\text{left}}(r) &=& e^{i k_{\text{in}} r}
	\mathbf{\chi}_\sigma + \sum_{\sigma^\prime}
	r_{\sigma^\prime\sigma} e^{-i k_{\text{in}} r}
	\mathbf{\chi}_{\sigma^\prime}\\ \Psi_{\text{right}}(r) &=&
	\sum_{\sigma^\prime} t_{\sigma^\prime\sigma} e^{i
	k_{\text{in}} r} \mathbf{\chi}_{\sigma^\prime},
\end{eqnarray} 
%
%
where $r$ is the coordinate on the semi-infinite input/output lead, 
with the origin fixed at the position of the input/output node.  

The transmission and reflection coefficients
($t_{\sigma^\prime\sigma}$ and $ r_{\sigma^\prime\sigma}$,
respectively) can be obtained by solving the linear system of
equations arising from the continuity of the probability current at
all nodes in the network and of the wavefunction at the input and
output nodes.  The conditions for the continuity of the probability
current at internal nodes are given in Eq.~(\ref{continuity}). For the
external nodes they read
%
%
\begin{eqnarray}
	\textbf{M}_{00} \mathbf{\Psi}_0 +\sum_{\langle 0 ,\beta
	\rangle} \textbf{M}_{0\beta}\mathbf{\Psi}_\beta &=&-
	i(\mathbf{\chi}_\sigma-\sum_{\sigma^\prime}
	r_{\sigma^\prime\sigma} \mathbf{\chi}_{\sigma^\prime})\\
	\textbf{M}_{NN} \mathbf{\Psi}_N +\sum_{\langle N ,\beta
	\rangle}\textbf{M}_{N\beta}\mathbf{\Psi}_\beta &=& i
	\sum_{\sigma^\prime} t_{\sigma^\prime\sigma}
	\mathbf{\chi}_{\sigma^\prime},
\end{eqnarray} 
%
%
where the injection node is labeled as ``$0$'' and the output node as
``$N$''.  The total transmission coefficient is then simply
$|t|^2=\sum_{\sigma,\sigma^\prime} |t_{\sigma^\prime\sigma}|^2$.
%
%
%
%
As it can be seen by inspection of the terms Eq.~(\ref{ms}) appearing
in the continuity equations (setting $l_{\alpha\beta}=L$), all the
properties are periodic in $k$ with a periodicity $2 \pi/L$.
Furthermore, for the total conductance the period in $k$ is halved,
i.e. it is $\pi/L$.  Finite temperature or finite voltage will
introduce in a natural way an average over $k_{\text{in}}$. For
$\mbox{Max}[{K_{\text{B}} T, eV}] \ge K_{\text{B}} T^{*}=
\frac{\hbar^2}{ m} k_F\frac {\pi}{L}$, the result of a transport
measurement will be the conductance integrated over $k_{\text{in}}\in
[0, \pi/L]$, indicated as $\langle G(k_{\text{SO}}\,L)
\rangle_{k_{\text{in}}}$. Taking for the Fermi energy of the
single-channel wires $10$ meV, $m/m_e=0.042$ for the effective mass
(InAs), and $L=1\mu$m, yields $T^{*}\approx 7$ K.
    
For a given $k_{\text{in}}$, the conductance has a rich structure that
takes into account the complexity of the associate energy spectrum. In
particular increasing $k_{\text{SO}}$ gaps open and the energy of the
incoming electrons ($\epsilon_{\text{in}}=\frac{\hbar^2
k_{\text{in}}^2}{2 m}$) can enter one of these gaps leading to a
vanishing conductance but not to localization [see panel b) of
Fig.~\ref{con_kf}].   
In fact, in this case the insulating behavior is
due to the absence of available states at the injection energy and not
to the localization in space of the electron wavefunction\cite{stubs}.
This effects is not present in $\langle G(k_{\text{SO}}\,L)
\rangle_{k_{\text{in}}}$, as the integration over $k_{\text{in}}$ is
equivalent to an average over energy.  The dependence of the average
conductance $\langle G(k_{\text{SO}}\,L)\rangle_{k_{\text{in}}}$ on
$k_{\text{SO}}$ is shown in panel a) of Fig. (\ref{con_kf}) for both
the diamond chain and the square ladder.  The conductance for both
kind of chains has a minimum for $k_{\text{SO}}\,L\,=\,\pi/2$ due to
interference caused by the phase differences induced by the Rashba
effect.  But due to the existence of the Rashba cages, this minimum
reaches zero only for the diamond chain.  In panel b) of
Fig~(\ref{con_kf}) the conductance for fixed $k_{\text{in}}$ for the
two chains is shown: for this choice of parameters, the gap opens only
for the diamond chain, while for the ladder a rich interference
pattern is present\cite{interference}.

\textit{Transport properties: disordered case.}
From the studies on the AB cages, we expect the localization induced
by the Rashba effect to be robust against disorder only in the
bipartite structure containing nodes with different coordination
numbers (diamond chain).  There are several kind of disorder that can
be considered.  Potential disorder along the wires (for example
randomly located point-like scatterers) does not lead, in this purely
one-dimensional model, to a modification of the phases acquired on a
bond by spin-precession but only to a renormalization of the bond
transmission.  The disorder that is more dangerous for the Rashba-cage
effect is a random fluctuation of the length of the bonds (see
Ref.~\cite{vidal-tra-2000}), as such length fluctuations induce
fluctuations of the phase shifts due to spin-precession.  Hence, we
consider a model where the length of each bond is randomly distributed
in the interval $[L-\Delta L,L+\Delta L]$.  The half width of the
distribution $\Delta L$ gives the strength of the disorder.
%
%
\begin{figure}
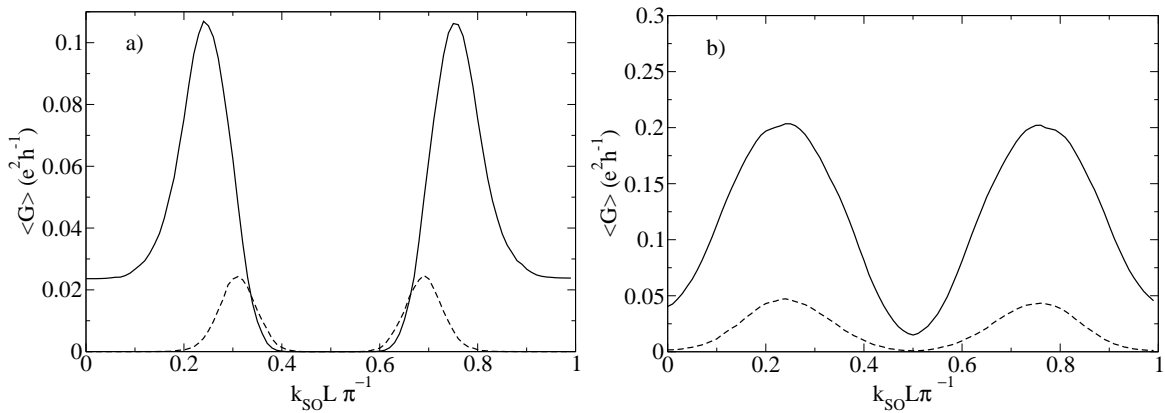

	\centering \includegraphics[width=3.in]{figure5.eps}
	\includegraphics[width=3.in]{figure6.eps} \caption
	{Conductance (averaged over disorder configurations and over
	$k_{\text{in}}$) plotted as a function of the spin-orbit
	coupling strength for the diamond chain [panel a)] and the 
	ladder [panel b)].  
        The two values of the disorder
	strength used in the calculation are: $\Delta L=0.01 L$ (solid
	line) and $\Delta L=0.02 L$ (dashed line). Disorder averaging
	is done over 50 configurations, and $k_\text{in}$ is uniformly
	distributed in $[k_{\text{F}}-\pi/2, k_{\text{F}}+\pi/2]$,
	with $k_{\text{F}} L=100$.  Both  
        systems 
        are composed by 50
	elementary loops.
\label{con_dis}}
\end{figure}
%
%
In order to clarify if disorder affects the conductance, we average
over disorder configurations.  This is relevant to experiments, as in
a real sample averaging is introduced by the finite phase-coherence
length.  For intermediate values of disorder ($k_{\text{F}} \Delta L
\approx 1$) we find that the Rashba-cage effect is still present for
the diamond chain, while the periodicity in $k_{\text{SO}}$ is halved
for the  
ladder,
as shown in Fig.~(\ref{con_dis}).  This latter
result can be interpreted as the analogous of the
Altshuler-Aharonov-Spivak (AAS) \cite{aas-1981} effect induced by the
SO coupling. The halving of the oscillation period is due to the
enhancement of back-reflection due to interference of pair of paths
traveling clockwise and counter-clockwise along a square of the chain
(according to weak localization picture).  At higher values of
disorder the AAS effect prevails also in the diamond chain. 

Finally, we expect that the results concerning the Rashba-cage effect 
will not change qualitatively when the wires are multimode if subband 
hybridization can be neglected, 
i.e. if the spin-precession length is much larger than the width of the 
wires. However, quantitative changes can occur due to scattering 
between the different modes at the vertices of the network.    
 
\textit{Conclusions.}
We have shown that in quantum network with a particular bipartite
geometry (diamond chain) is possible to obtain localization of the
electron wavefunction by means of the Rashba effect.  This
localization shows up both in the spectrum of the infinite chain which
becomes nondispersive, and in the transport properties of a
finite-size chain connected to leads.  Furthermore, transport
calculations in the presence of disorder show that in bipartite
structure containing nodes with different connectivity this
Rashba-cage effect is robust against disorder.

We gratefully acknowledge helpful discussions with 
 G. De Filippis, R. Fazio, D. Frustaglia and  A.C. Perroni.


\begin{thebibliography}{99}
\bibitem{vidal-1998} J. Vidal, R. Mosseri, and B. Dou\c cot,
Phys. Rev. Lett. {\bf 81},   5888 (1998).

\bibitem{vidal-cage-2000} J. Vidal, B. Dou\c cot, R. Mosseri, and
P. Butaud, Phys. Rev. Lett. \textbf{85}, 3906 (2000).

\bibitem{anderson-1958} P.W. Anderson, Phys. Rev. \textbf{109}, 1492 (1958).

\bibitem{vidal-prb-long}  J. Vidal, P. Butaud, B. Dou\c cot, and
R. Mosseri, Phys. Rev. B \textbf{64}, 155306 (2001)

\bibitem{vidal-tra-2000} J. Vidal, G. Montambaux, and B. Dou\c cot,
Phys. Rev. B  \textbf{62}, R16294 (2000).

\bibitem{abilio-1999} C.C. Abilio, P. Butaud, T. Fournier,
B. Pannetier, J. Vidal, S. Tedesco, and B. Dalzotto,
Phys. Rev. Lett. \textbf{83}, 5102 (1999).

\bibitem{naud-2001} C. Naud, G. Faini, and D. Mailly,
Phys. Rev. Lett. \textbf{86}, 5104 (2001).

\bibitem{casher} Y. Aharonov and A. Casher,
Phys. Rev. Lett. \textbf{53}, 319 (1984).


\bibitem{mathur} H. Mathur, and A. Douglas Stone, Phys. Rev. Lett. 
{\textbf 68}, 2964 (1992)

\bibitem{balatsky} A.V. Balatsky, and B. L. Altshuler, Phys. Rev. Lett.  
\textbf{70} 1678 (1993).


\bibitem{aronov-1993} A.~G. Aronov and Y.~B. Lyanda-Geller,
Phys. Rev. Lett. \textbf{70}, 343 (1993).

\bibitem{splettstoesser-2003} J. Splettst\"o{\ss}er, M. Governale,
and U. Z\" ulicke, Phys. Rev. B \textbf{68}, 165341 (2003).

\bibitem{frustaglia} D. Frustaglia and K. Richter, cond-mat/0309228.

\bibitem{rashba-1960} E. Rashba, Fiz. Tverd. Tela (Leningrad)
\textbf{2}, 1224 (1960), [Sov. Phys. Solid State 2, 1109 (1960)].

\bibitem{nitta-1997} J. Nitta, T. Akazaki, H. Takayanagi, and
T. Enoki, Phys. Rev. Lett. \textbf{78}, 1335 (1997).

\bibitem{schaepers-1998} T. Sch\" apers, J. Engels, T. Klocke,
M. Hollfelder, and H. L\" uth, J. Appl. Phys. \textbf{83}, 4324
(1998). 

\bibitem{grundler-2000} D. Grundler, Phys. Rev. Lett. \textbf{84},
6074 (2000).

\bibitem{mireles-2001} F. Mireles and G. Kirczenow,
Phys. Rev. B \textbf{64}, 24426 (2001).

\bibitem{governale-2002} M. Governale and U. Z\" ulicke,
Phys. Rev. B \textbf{66}, 073311 (2002).

\bibitem{note2}The term
in $k_{\text{SO}}^2$ can be neglected in realistic situations.

\bibitem{kottos-1999} T. Kottos and U. Smilansky, Ann. Phys. (N.Y.)
\textbf{274}, 76 (1999).

\bibitem{landauer-1988} R. Landauer, IBM J. Res. Dev. \textbf{1}, 223 (1957).

\bibitem{buttiker-1988} M. Buttiker, IBM J. Res. Dev. \textbf{32}, 317 (1988).

\bibitem{stubs} Conductance gaps have been found in quantum wires with 
Rashba spin-orbit coupling, in the presence of a periodic array of 
stubs, as a function of the stub length or width 
in X. F, Wang, P. Vassiolopoulos, and F. M. Peeters,
Appl. Phys. Lett. {\textbf{80}}, 1400 (2002).  

\bibitem{interference} The complexity of the interference pattern 
is related to the many different interfering 
paths allowed by the ladder
geometry. 

\bibitem{aas-1981} B. Altshuler, A. Aharonov, and B. Spivak, Pis'ma
Zh. Eksp. Teor. Fiz. \textbf{33}, 101 (1981), [JEPT Lett.
  \textbf{33}, 94, (1981)].

\end{thebibliography}
\end{document}